\documentclass[twocolumn,showpacs,preprintnumbers,amsmath,amssymb, superscriptaddress, prl]{revtex4}

\usepackage{graphicx}
\usepackage{dcolumn}
\usepackage{bm}
\usepackage{color}
\usepackage{textcomp}
\definecolor{red}{rgb}{1,0,0}

\definecolor{green}{rgb}{0,1,0}

\definecolor{blue}{rgb}{0,0,1}

\begin{document}

\title{Spin-stripe phase in a frustrated zigzag spin -$\frac{1}{2}$ chain}
\author{M. Pregelj}
\email{matej.pregelj@ijs.si}
\affiliation{Jo\v{z}ef Stefan, Jamova c. 39, 1000 Ljubljana, Slovenia}
\author{A. Zorko}
\affiliation{Jo\v{z}ef Stefan, Jamova c. 39, 1000 Ljubljana, Slovenia}
\author{O. Zaharko}
\affiliation{Laboratory for Neutron Scattering and Imaging, Paul Scherrer Institute, CH-5232 Villigen, Switzerland}
\author{H. Nojiri}
\affiliation{Institute for Materials Research, Tohoku University, Sendai 980-8577, Japan}
\author{H. Berger}
\affiliation{Ecole polytechnique f\'{e}d\'{e}rale de Lausanne, CH-1015 Lausanne, Switzerland}
\author{L. C. Chapon}
\affiliation{Institut Laue-Langevin, BP 156X, 38042 Grenoble, France}
\author{D. Ar\v{c}on}
\affiliation{Jo\v{z}ef Stefan, Jamova c. 39, 1000 Ljubljana, Slovenia}
\affiliation{Faculty of Mathematics and Physics, University of Ljubljana, Jadranska c. 19, 1000 Ljubljana, Slovenia}
\date{\today}

\begin{abstract}

Motifs of periodic modulations are encountered in a variety of natural systems, where at least two rival states are present.
In strongly correlated electron systems such behaviour has typically been associated with competition between short- and long-range interactions, e.g., between exchange and dipole-dipole interactions in the case of ferromagnetic thin films. 
Here we show that spin-stripe textures may develop also in antiferromagnets, where long-range dipole-dipole magnetic interactions are absent.
A comprehensive analysis of magnetic susceptibility, high-field magnetization, specific heat, and neutron diffraction measurements unveils $\beta$-TeVO$_4$ as a nearly perfect realization of a frustrated (zigzag) ferromagnetic spin-1/2 chain.
Strikingly, a narrow spin stripe phase develops at elevated magnetic fields due to weak frustrated short-range interchain exchange interactions possibly assisted by the symmetry allowed electric polarization.
This concept provides an alternative route for the stripe formation in strongly correlated electron systems and may help understanding other widespread, yet still elusive, stripe-related phenomena.

\end{abstract}

\maketitle

Stripe patterns appear in remarkably diverse materials, ranging from biological to strongly-correlated electron systems \cite{Seul, Stoycheva, Malescio, DeBell, Portmann, Mu, Giuliani, Andelman, Seifert, Baumgart, Tranquada, Emery, Vojta, Ghiringhelli, Wu2013}.
The common driving force behind stripe formation is a natural tendency of a system to balance two or more rival phases that appear due to competing interactions.
For instance, in ferromagnetic thin films spin-up and spin-down magnetic domains form stripe textures due to opposing short-range exchange and long-range dipole-dipole interactions \cite{DeBell, Portmann, Mu, Giuliani}.
In antiferromagnets, however, there is no finite magnetization that could mediate long-range interactions between magnetic domains, so the stripe formation is not expected.
Nevertheless, stripe modulation may develop also due to the coupling between different order parameters that is mediated solely by short-range forces, as has been observed in some biological membrane systems \cite{Andelman,Seifert,Baumgart}.
Similar mechanism can be active also in antiferromagnets and may be relevant to other strongly correlated electron systems, e.g., to high-temperature superconductors, where the competition between charge-stripe ordering \cite{Tranquada, Emery, Vojta} and superconductivity has become increasingly apparent \cite{Ghiringhelli, Wu2013}.
Therefore, any evidence of stripe formation in the absence of long-range interactions in strongly correlated electron systems is of utmost importance.

\begin{figure} [!] \center
\includegraphics[trim = 0mm 0mm 0mm 0mm, clip, width=0.46\textwidth]{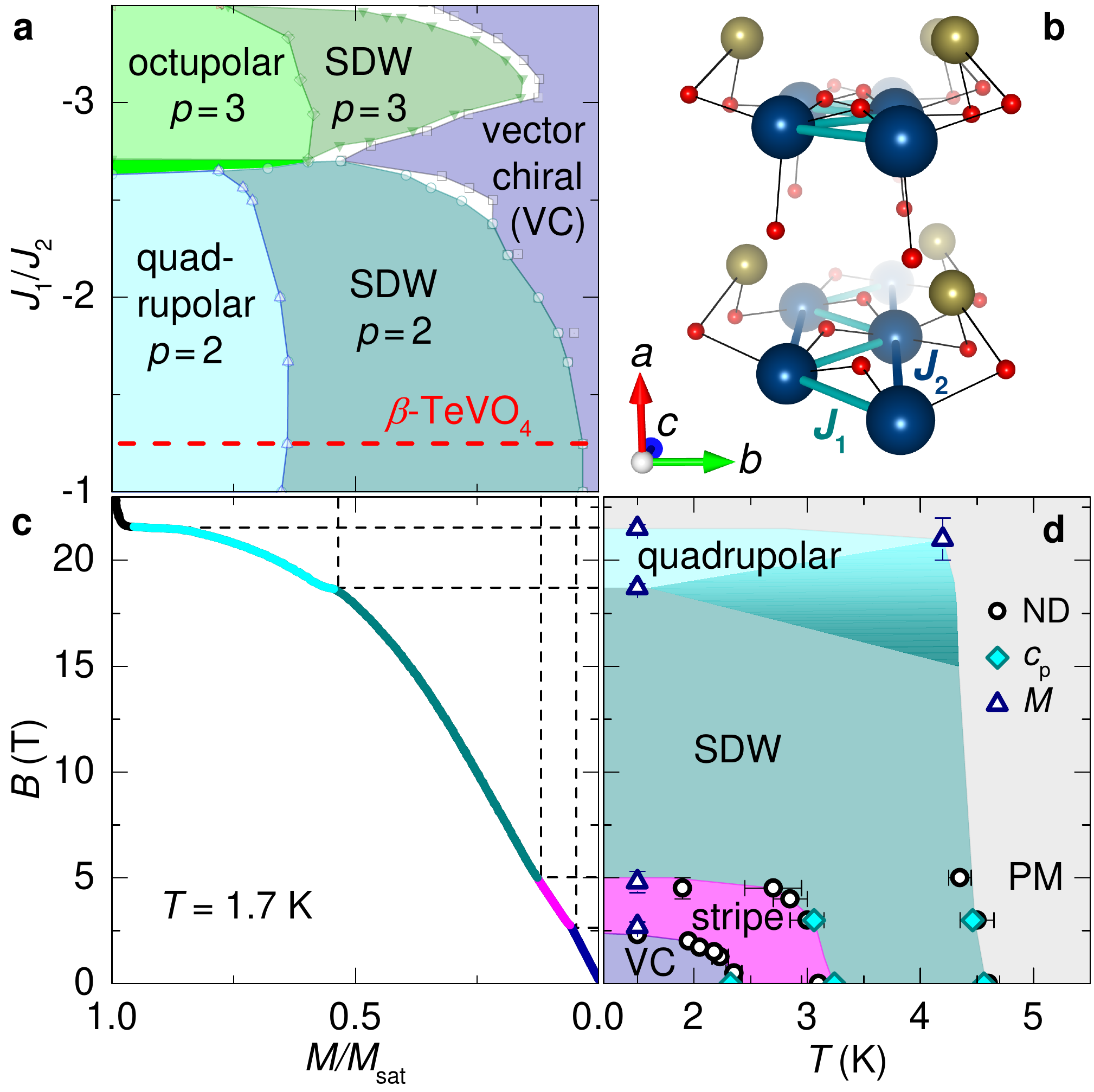}
\caption{{\bf Comparison of the theoretical and experimental phase diagrams.} ({\bf a}) Schematic phase diagram of the frustrated ferromagnetic spin-1/2 chain model as a function of $J_1 /J_2$ and $M/M_{\text{sat}}$ (for details see \cite{Sudan}), where VC and SDW denote vector-chiral and spin-density-wave phases, respectively, while $p$ denotes the order of the bound magnon state. The dashed line corresponds to $J_1 /J_2$\,=\,$-1.25$, found in $\beta$-TeVO$_4$. ({\bf b}) The crystal structure of $\beta$-TeVO$_4$, with zigzag-chain interactions. Small, medium and large spheres denote O$^{2-}$, Te$^{4+}$ and the magnetic V$^{4+}$ ions, respectively. ({\bf c}) Magnetization normalized to the saturation value ($M/M_{\text{sat}}$) measured in the magnetic field along the $a$ axis and ({\bf d}) the experimental magnetic phase diagram of $\beta$-TeVO$_4$, with PM indicating the paramagnetic state, while ND,  $c_\text{p}$ and $M$ denote data points derived from neutron diffraction, specific heat and magnetization measurements, respectively. Error bars denote the uncertainty of the magnetic transitions.}
\label{PD}
\end{figure}

A convenient system to search for this intriguing phenomenon is a frustrated (zigzag) quantum spin-1/2 chain, because of its extremely rich phase diagram. 
Most importantly, the energy difference between collinear and spiral spin states can be accurately tuned by the applied magnetic field, $B$, \cite{Sudan, Hikihara, Hikihara2010, Okunishi, McCulloch}.
In particular, when the nearest-neighbour  (NN) interaction ($J_1$) is ferromagnetic and the next-nearest-neighbour  (NNN) one is antiferromagnetic ($J_2$\,$>$\,0), i.e., in the case of a frustrated ferromagnetic zigzag chain, a vector-chiral (VC) ground state is at elevated magnetic fields succeeded by collinear spin-density-wave (SDW) phases \cite{Sudan, Hikihara}, with two- ($p$\,=\,2) or three-magnon ($p$\,=\,3) bound states (Fig.\,\ref{PD}a).
Moreover, when magnetization $M$ approaches saturation ($M/M_{\text{sat}}$\,=\,1), intriguing multipolar orders develop \cite{Sudan, Hikihara}.
Although several coper-oxide \cite{Schapers,Sato13} and vanadate \cite{Chakrabarty, Kataev, Svistov} compounds possess the "zigzag" chain topology, high saturation fields, anisotropy effects, or sizable interchain interactions hamper experimental investigation in most cases.

Here we highlight $\beta$-TeVO$_4$ where chains of distorted corner-sharing VO$_5$ pyramids run along the $c$ axis \cite{Savina, Meunier2}.
So far, this compound has been treated as a simple V$^{4+}$ spin-1/2 chain system with only NN antiferromagentic interaction [$J_1$\,=\,21.4(2)\,K] \cite{Savina}, as the maximum in the magnetic susceptibility, $\chi(T)$, at 14\,K can be described by the Bonner-Fischer model \cite{Bonner}.
However, high-temperature susceptibility yields a very small positive (ferromagnetic) Curie-Weiss temperature $\theta$, varying from 1.6 to 4.4\,K, depending on the magnetic field orientation \cite{Savina}, which
implies the importance of both the NN V$-$O$-$V and the NNN V$-$O$-$Te$-$O$-$V exchange interactions (Fig.\,\ref{PD}b) and their opposite sign.
Finally, three subsequent magnetic transitions were reported at $T_{\text{N1}}$\,=\,4.65\,K, $T_{\text{N2}}$\,=\,3.28\,K, and $T_{\text{N3}}$\,=\,2.28\,K \cite{Savina}, signifying several energetically almost equivalent magnetic states -- a signature of competing magnetic interactions and strong frustration.

In this work we present magnetic susceptibility, magnetization, specific heat, neutron diffraction, and spherical neutron polarimetry measurements, which reveal that competing ferromagnetic NN and antiferromagnetic NNN interactions in $\beta$-TeVO$_4$ are comparable in magnitude.
The derived magnetic phase diagram corroborates main features of the theoretical one (Fig.\,\ref{PD}), thus making $\beta$-TeVO$_4$ a nearly ideal realization of the frustrated ferromagnetic spin-1/2 chain model, where quadrupolar/spin-nematic phase is predicted already at 18.7\,T.
Moreover, a striking new stripe phase has been discovered between the theoretically-predicted VC and SDW phases.
In the corresponding magnetic field range, the strength of the interchain interaction matches the energy difference between these two parent phases and thus allows for the establishment of the nanoscale modulation of the magnetic order even in the absence of sizable long-range dipole-dipole magnetic interactions.
Possible electric polarization that is in the VC state allowed by symmetry may further assist the stripe formation.
This makes $\beta$-TeVO$_4$ an intriguing manifestation of the spin-stripe modulation in an antiferromagnet, where stripe formation appears to be driven by the coupling between order parameters, a mechanism so far almost exclusively attributed to biological systems.

\section{Results}

\begin{figure} [!] \center
\includegraphics[trim = 2mm 0mm 0mm 0mm, clip, width=0.48\textwidth]{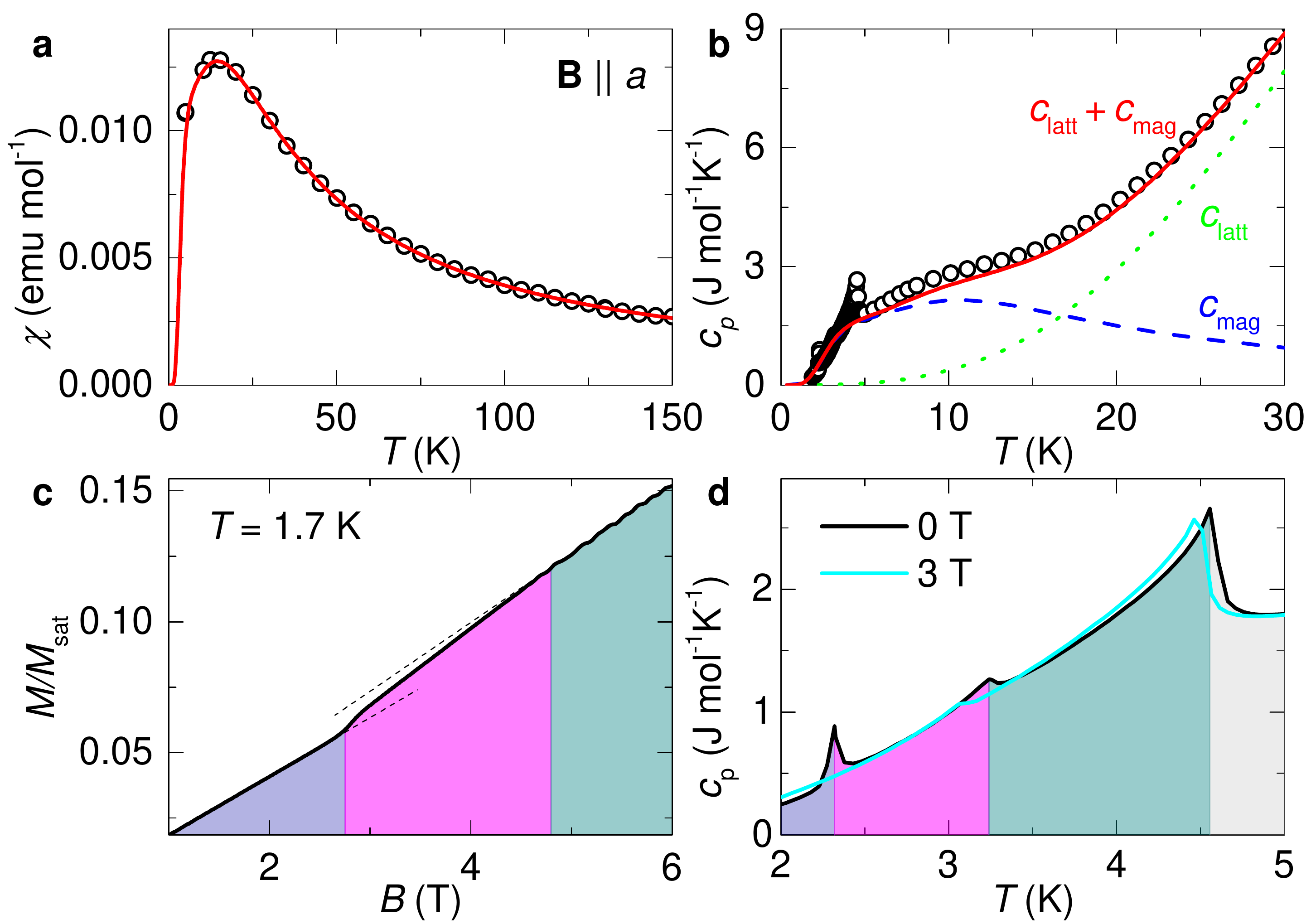}
\caption{{\bf Bulk measurements.} ({\bf a}) Magnetic susceptibility (symbols) at 0.1\,T ({\bf B}$||a$) and the fit to the $J_1/J_2$ zigzag chain model assuming an average $z'J'$ interchain coupling (line).
({\bf b}) Specific heat in zero-field (symbols) and the fit ($c_{\text{latt}}$+$c_{\text{mag}}$; solid line), considering the lattice ($c_{\text{latt}}$; dotted) and the magnetic ($c_{\text{mag}}$; dashed) contributions.
({\bf c}) The low-field part of the magnetization for {\bf B}$||a$. The dashed lines are linear extrapolations, indicating two magnetic transitions.
({\bf d}) The low-temperature specific heat measured at 0 and 3\,T ({\bf B}$||a$).
The colour coding denotes the VC (purple), stripe (magenta) and SDW (dark cyan) ordered phases and the disordered paramagnetic phase (grey). }
\label{bulk}
\end{figure}
{\bf Signatures of multiple magnetic transitions.}
In agreement with the published results \cite{Savina}, our magnetic susceptibility data $\chi(T)$ above 80\,K for {\bf B}$||a$ yield ferromagnetic character [$\theta$\,=\,2.7(1)\,K; see Methods] as well as a clear maximum at much higher temperature of 14(1)\,K (Fig.\,\ref{bulk}a).
Similarly, specific-heat results show a pronounced hump at 12(1)\,K (Fig.\,\ref{bulk}b), which corresponds to the magnetic part of the specific heat $c_{\text{mag}}$ that develops on top of the usual lattice contribution $c_{\rm{latt}}$ \cite{Debye} (see Methods), as expected for a low-dimensional system.
At lower temperatures, distinct $\lambda$-type anomalies are observed at $T_{\text{N1}}$ and $T_{\text{N3}}$, whereas the kink at $T_{\text{N2}}$ is more subtle (Fig.\,\ref{bulk}d).
When the magnetic field of 3\,T is applied along the $a$ axis, the anomaly at $T_{\text{N3}}$ disappears, implying that the zero-field magnetic ground state is suppressed.
\begin{figure*} [!] 
\includegraphics[trim = 0mm 0mm 0mm 0mm, clip, width=0.85\textwidth]{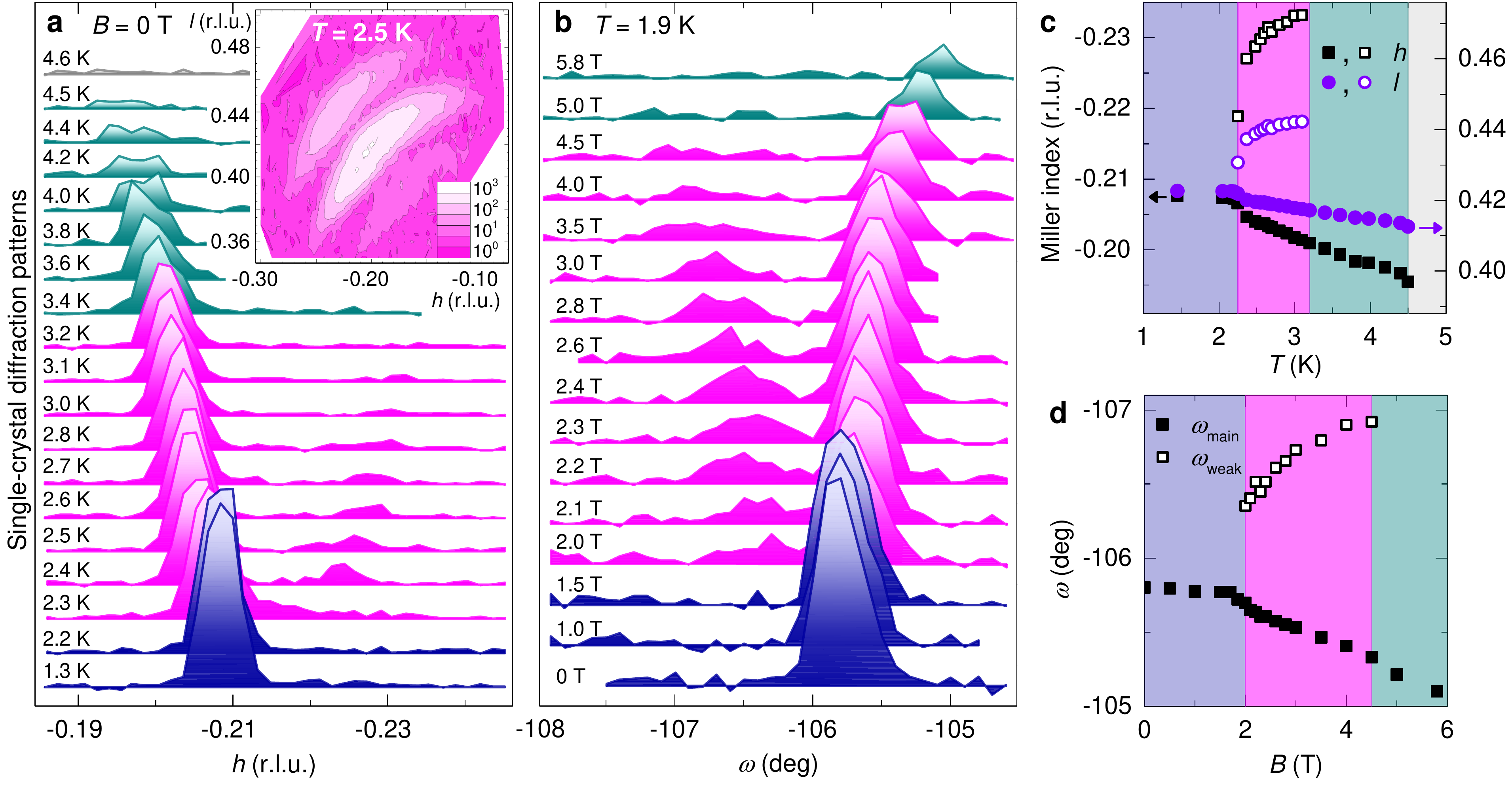}
\caption{{\bf Neutron diffraction results.} ({\bf a}) The temperature evolution of the main ($-0.208$~0~0.423) magnetic reflection. Inset: the corresponding {\bf k}-map measured at 2.5\,K. ({\bf b}) The magnetic-field dependence ({\bf B}$||a$) of the (0.208~2~0.577) reflection at 1.9\,K and the derived ({\bf c}) temperature- and ({\bf d}) field-dependent position of the main (solid symbols) and weak (open symbols) magnetic reflections. The colour coding denotes the VC (purple), stripe (magenta) and SDW (dark cyan) ordered phases and the disordered paramagnetic phase (grey).}
\label{ND}
\end{figure*}
This corroborates with the magnetization measured for the same orientation of the magnetic field at 1.7\,K, i.e, well below $T_{\text{N3}}$, which exhibits two distinct anomalies at $B_{\text{c1}}$\,=\,2.8\,T (Fig.\,\ref{bulk}c) and at $B_{\text{c3}}$\,=\,18.7\,T and finally reaches full saturation at $B_{\text{sat}}$\,=\,21.5\,T (Fig.\,\ref{PD}c).
A detailed inspection reveals an additional inflection point at $B_{\text{c2}}$\,=\,4.5(4)\,T (Fig.\,\ref{bulk}c).
Evidently, below $T_{\text{N3}}$, $\beta$-TeVO$_4$ undergoes three magnetic-field-induced transitions before reaching full magnetization saturation.

{\bf Incommensurability revealed by neutron diffraction.}
Neutron experiments disclose a zero-field spiral magnetic ground state existing below $T_{\text{N3}}$ that can be described with a single magnetic wave vector {\bf k}\,=\,($-0.208$~0~0.423) (Fig.\,\ref{ND}a) (see Methods).
The observed incommensurability is a clear signature of the magnetic frustration.
Therefore, the modelling of $\chi(T)$ with the Bonner-Fisher model of the simple NN spin-1/2 chain \cite{Savina} is not appropriate.
Considering the zigzag-chain structure (Fig.\,\ref{PD}b), the frustration in $\beta$-TeVO$_4$ must originate from the competition between the NN and the NNN exchange interactions, $J_1$ and $J_2$, respectively.
Indeed, the NN bond is similar to the corner-sharing V$-$O$-$V  bridge in CdVO$_3$ \cite{Tsirlin} and may carry ferromagnetic exchange \cite{Dai}.
The NNN bridge, on the other hand, involves the Te$^{4+}$ lone-pair cation, capable of bridging sizable antiferromagnetic exchange interactions \cite{FTOBAFMR}. 
The pitch angle between two NN spins in a zigzag chain is classically defined as $\phi$\,=\,arccos($-J_1/4J_2$) \cite{Bursill, White}, reflecting in $k$\,=\,$\phi/2\pi$.
However, since the two NN V$^{4+}$ ions in $\beta$-TeVO$_4$ lie in the same unit cell, i.e., $k_z$ associates NNN spins, the experimentally determined $k_z$\,=\,0.423 must be scaled by 1/2, which yields the NN pitch angle $\phi_c$\,=\,0.423\,$\pi$ and thus $|J_1/J_2|$\,$\sim$\,1.
Taking into the account the experimentally determined Curie-Weiss temperature $\theta$\,=\,2.7(1)\,K (Methods), which is even smaller than the ordering temperature $T_{\text{N1}}$\,=\,4.65\,K, the competing $J_1$  and $J_2$ interactions must have different signs ($J_1/J_2$\,$\sim$\,$-$1).
In fact, to assure the competition $J_1$ must be ferromagnetic and $J_2$ must be antiferromagnetic.
In contrast to the chain ($c$) direction, the two neighbouring spins along the $a$ axis lie in the neighbouring unit cells.
The experimentally derived $k_z/k_x$\,$\sim$\,$-2$ thus yields a similar pitch angle along the $a$ axis, $\phi_a$\,$\sim$\,$-\phi_c$.
The V$-$O interchain distance [$d$\,$=$\,2.81(5)\,\AA\,, see Methods] is much larger than the typical V$-$O exchange leg in vanadates, $\sim$2.0(1)\,\AA\, \cite{Koo, Onoda, Pickett}, hence, the interchain exchange most likely proceeds via Te$^{4+}$ ions, as frequently found in oxyhalide tellurites \cite{FTOBAFMR}.
In this respect, two very similar interchain V$-$O$-$Te$-$O$-$V exchange paths exist (see Methods), suggesting that the interchain interactions are also frustrated and that the exact relation between $\phi_a$ and $\phi_c$ is nontrivial.

{\bf Quantitative determination of the exchange interactions.}
In order to place our system in the general zigzag spin-chain phase diagram (Fig.\,\ref{PD}a) \cite{Sudan, Hikihara, Hikihara2010}, we still need to determine the magnitude of the exchange interactions.
This can be determined from modeling of $\chi(T)$.
We fit $\chi(T)$ using the exact-diagonalization results for the isolated zigzag chain $\chi_{\text{zig}}(T)$ \cite{Chakrabarty, Heidrich-Meisner, Heidrich-Meisner-online} with an additional interchain exchange within a mean-field approach  \cite{Carlin, Chakrabarty}, $\chi_{\text{coupled}}(T)$\,=\,$\chi_{\text{zig}}/(1-\lambda \chi_{\text{zig}})$.
Here $\lambda\,=\,z'J'/(N_\text{A} g^2\mu_\text{B}^2)$, $J'$ is the interchain coupling to $z'$ spins on the neighbouring chains, $N_\text{A}$ is the Avogadro constant, $\mu_\text{B}$ is the Bohr magneton, and $g$\,=\,2.01(1) is the gyromagnetic factor for V$^{4+}$ (see Methods).
An excellent agreement with the experiment is obtained for $J_1/J_2$\,=\,$-1.25$, with $J_1$\,=\,38.34\,K and $z'J'$\,=\,4.46\,K (Fig.\,\ref{bulk}a), which is in good agreement with the classical estimate $J_1/J_2$\,$\sim$\,$-$1 derived from the experimentally determined $k_z$.
We note that the increased $|J_1/J_2|$ value is an expected effect of quantum fluctuations for spin-1/2 systems \cite{Bursill}.
On the other hand, fitting with $J_1/J_2$\,$>$\,0 yields unrealistically large $|z'J'|$\,$>$\,$|J_1|, |J_2|$.
Moreover, the calculated Curie-Weiss temperature \cite{Kittel} $\theta$\,=\,$-\frac{2}{3}S(S$+1)$\sum_iz_iJ_i$\,= \,$-\frac{1}{2}$[2($J_1$+$J_2$)+$z'J'$]\,=\,5.4\,K is in fair agreement with the experimental value $\theta$\,=\,2.7(1)\,K, considering an order of magnitude larger exchange interactions. 
Finally, using $c_{\rm{mag}}(T)$ \cite{Heidrich-Meisner,Heidrich-Meisner-online} for the same $J_1$, $J_2$ and adding it to the lattice contribution $c_{\rm{latt}}(T)$ within the Debye model (see Methods), we obtain an excellent fit of the specific heat data (Fig.\,\ref{bulk}b), thus affirming the extracted exchange parameters.

\begin{figure*} [!]
\includegraphics[trim = 0mm 60mm 0mm 0mm, clip, width=0.98\textwidth]{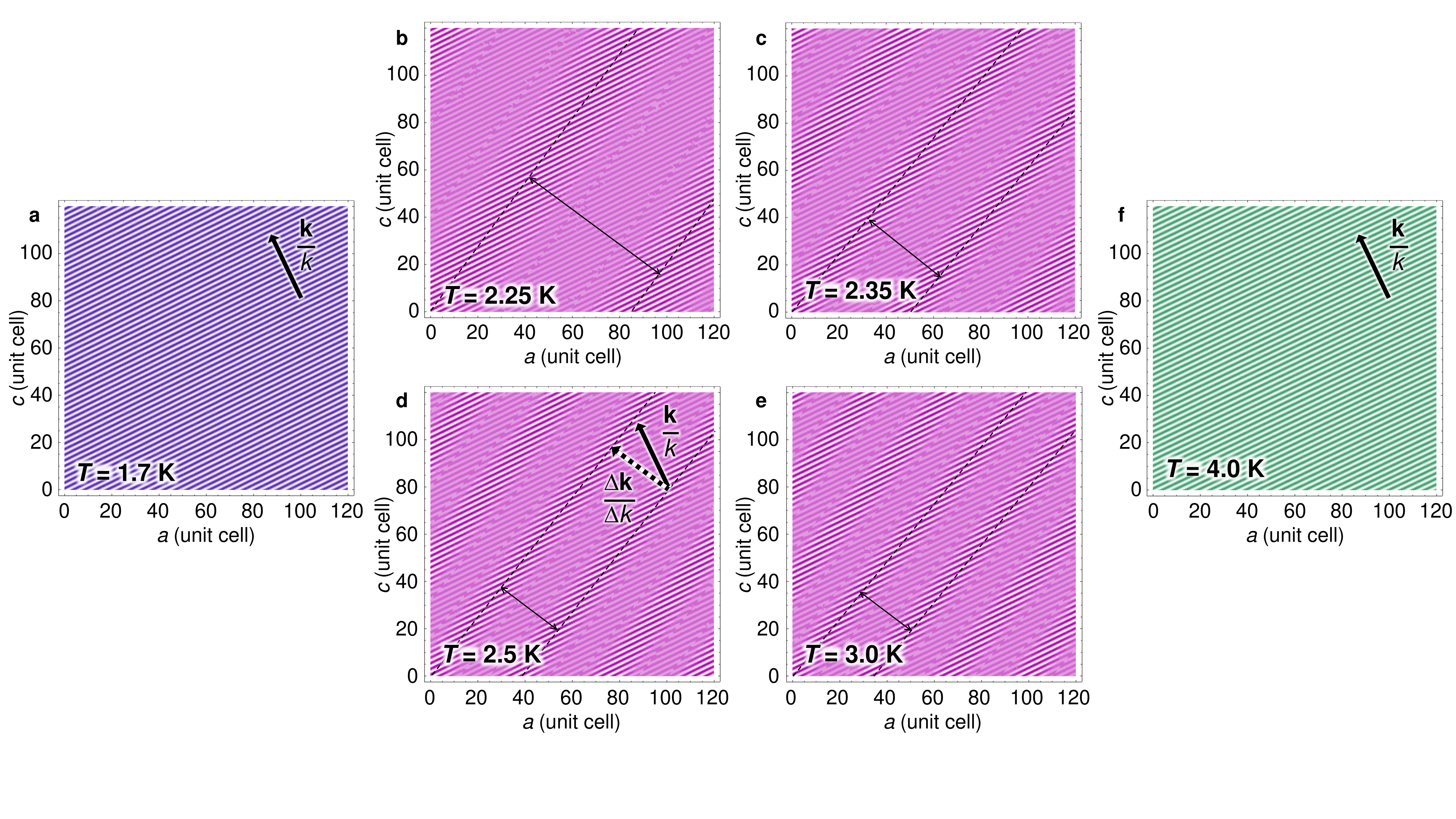}
\caption{{\bf Temperature evolution of the magnetic structure modulation.} The zero-field magnetic structure modulation ({\bf a}) in the VC phase at 1.7\,K, ({\bf b}-{\bf e}) in the stripe phase between 2.25 and 3.0\,K, where super-satellite reflections reveal additional nanometer-scale modulation, and ({\bf f}) in the SDW phase at 4.0\,K, where super-satellite reflections are absent again.
Different magnetic phases, i.e., the VC (purple), stripe (magenta) and SDW (dark cyan) phases, are indicated by different colours. The colour scale denotes the magnetic modulation, the direction of which, {\bf k}/$k$, is indicated by thick arrows. In the stripe phase, the direction of the additional stripe modulation, $\Delta${\bf k}/$\Delta k$, is indicated by the thick dotted arrow. The thin dashed lines emphasize the centres of the stripes, allowing for the estimate of the stripe modulation period (two sided arrows), which amounts to ({\bf b}) 30.7\,nm, ({\bf c}) 17.8\,nm, ({\bf d}) 12.7\,nm and ({\bf e}) 12.3\,nm.}
\label{mag-struc}
\end{figure*}

{\bf Determination of the magnetic phase diagram.}
To explore all magnetic phases of $\beta$-TeVO$_4$ we performed detailed temperature- and magnetic-field-dependent neutron diffraction experiments, focusing on the strongest magnetic reflection (Fig.\,\ref{ND}).
In zero field, the reflection emerges at $T_{\text{N1}}$\,=\,4.5(1)\,K at {\bf k}\,=\,($-0.195$~0~0.413) and shifts to larger $|h|$ and $|l|$ with decreasing temperature.
At $T_{\text{N2}}$\,=\,3.1(1)\,K, the main {\bf k} reflection is accompanied by an additional weak reflection appearing at {\bf k'}\,=\,($-0.233$~0~0.442).
On further cooling, this reflection shifts towards the main one and at $T_{\text{N3}}$\,=\,2.20(3)\,K they finally collapse into a single peak at {\bf k}\,=\,($-0.208$~0~0.423).
This peak does not shift with temperature at least down to 1.6\,K (Fig.\,\ref{ND}c), indicating the stability of the magnetic ground state.
Very similar response is observed below $T_{\text{N3}}$ when magnetic field is applied along the $a$ axis.
Above $B_{\text{c1}}$, the weak {\bf k'} reflection separates from the main {\bf k} reflection and eventually disappears at $B$\,$\sim$\,$B_{\text{c2}}$ (Fig.\,\ref{ND}b).
It thus appears that both external controlling parameters, i.e., the temperature and the applied magnetic field, drive the system through the same magnetic phases.

Further details of the three magnetic phases are provided by zero-field spherical neutron polarimetry. In our study we focused on the strongest magnetic ($-0.208$~0~0.423) reflection (see Methods).
In particular, we measured the full polarization matrix at 1.6\,K and then followed the temperature evolution of chiral $yx$ and non-chiral $yy$ terms (Table\,\ref{tabSNP}).
Measurements in the ($h$00)/(00$l$) scattering plane reveal a significant chiral term below $T_{\text{N3}}$, indicating an imbalance in the chiral-domain population.
Above $T_{\text{N3}}$, on the contrary, the chiral term is almost completely suppressed, implying that chirality is either dramatically reduced or the plane of the spiral is reoriented along the scattering plane.
This complies with the behavior of the non-chiral term, which above $T_{\text{N3}}$ significantly increases, indicating a reorientation of the magnetic moments.
Measuring the same reflection in the ($-h$02$h$)/(0$k$0) scattering plane allows probing a different projection of the magnetic structure factor. 
We find that the chiral term is now absent in the entire temperature range (Table\,\ref{tabSNP}), while the non-chiral term again increases above $T_{\text{N3}}$.
We can thus conclude that (i) the plane of the spiral, existing below  $T_{\text{N3}}$, is very close to the ($-h$02$h$)/(0$k$0) scattering plane, (ii) at $T_{\text{N3}}$ magnetic moments reorient and (iii) above $T_{\text{N3}}$ magnetic order remains incommensurate, but is dominated by amplitude modulation (SDW).

\begin{table} [!]
\caption{{\bf Elements of the polarization matrix.} Chiral ($yx$) and non-chiral ($yy$) terms of the polarization matrix of the ($-0.208$~0~0.423) magnetic reflection, measured in two different scattering planes; S1 ($h$00)/(00$l$) and S2 ($-h$02$h$)/(0$k$0) and corrected for the incomplete polarization of incoming neutrons.
\label{tabSNP}}
\begin{ruledtabular}
\begin{tabular}{ccccc} 
 $T$ (K)   &  \multicolumn{2}{c}{$yx$} & \multicolumn{2}{c}{$yy$}\\
\hline
      & S1       &     S2    & S1       &     S2    \\
1.6 & 0.33(4) & 0.07(5) & 0.34(4) & -0.27(5) \\
2.5 & 0.11(5) & 0.04(7) & 1.01(4) & -0.66(7) \\
3.5 & 0.04(6) &         & 0.95(4) &          \\
\end{tabular}
\end{ruledtabular}
\end{table}

All our data are summarized in the magnetic phase diagram (Fig.\,\ref{PD}d), which we compare to the theoretical diagram for the frustrated ferromagnetic ($J_1/J_2$\,$<$\,$0$) zigzag spin-1/2 chain model in Fig.\,\ref{PD}.
This comparison is best presented when the low-temperature magnetization data (Fig.\,\ref{PD}c) are used for scaling between $B$ and $M/M_{\text{sat}}$.
The zero-field spiral magnetic ground state clearly confirms the predicted VC state.
Above $B_{\text{c2}}$, the intensity of the magnetic reflection significantly changes (Fig.\,\ref{ND}b), while the magnetic order remains incommensurate.
Hence, the remarkable agreement between the theoretical and experimental $M/M_{\text{sat}}$ values at $B_{\text{c2}}$ and $B_{\text{c3}}$ (Fig.\,\ref{PD}) suggests that the VC phase ($B$\,$<$\,$B_{\text{c1}}$) is followed by the SDW phase ($B_{\text{c2}}$\,$<$\,$B$\,$<$\,$B_{\text{c3}}$).
This complies with spherical neutron polarimetry results, indicating the presence of SDW above $T_{\text{N3}}$.
The SDW phase is according to the model succeeded by the quadrupolar/spin-nematic phase persisting up to the saturation ($B_{\text{c3}}$\,$<$\,$B$\,$<$\,$B_{\text{sat}}$).
We note that the steep magnetization slope in this region differs from the gentle slope experimentally observed in LiCuVO$_4$ \cite{Svistov}, presumably due to the different interchain couplings in the two systems.
However, the narrow phase between $B_{\text{c1}}$ and $B_{\text{c2}}$, where the weak {\bf k'} reflection complements the main {\bf k} magnetic reflection, is observed for the first time.

{\bf Emergence of a novel spin-stripe phase.}
To gain more information about the peculiar {\bf k'} reflection, we measured an extended {\bf k}-map at 2.5\,K, i.e., between $T_{\text{N2}}$ and $T_{\text{N3}}$  (inset in Fig.\,\ref{ND}a).
Evidently, the main magnetic reflection at {\bf k}(2.5\,K)\,=\,($-0.203$~0~0.419) is accompanied with two ``super-satellite'' reflections symmetrically shifted by $\pm\Delta${\bf k}(2.5\,K)\,=\, $\pm$($-0.030$~0~0.021) from {\bf k}(2.5\,K).
This indicates that the incommensurate magnetic structure in this phase experiences an additional nanometer-scale stripe modulation with well-defined modulation period of 12.7\,nm, i.e., $\sim$40 lattice units (Fig.\,\ref{mag-struc}d). 
Moreover, by heating from 2.25 to 3.1\,K, this modulation period can be changed between 30 to 12\,nm, respectively (Fig.\,\ref{mag-struc}b-\ref{mag-struc}e).
We stress that the width of all magnetic reflections, i.e., the main as well as satellite ones, is resolution limited, indicating that both magnetic modulations ({\bf k} and $\Delta${\bf k}) are very well defined, which is a clear evidence of a coherent and uniform magnetic state rather than a simple coexistence of the VC and the SDW phases.
The observed $\Delta${\bf k} modulation markedly differs from typical solition lattices, which develop in the insulating SDW phases at low temperatures.
There the sinusoidal modulation squares up, reflecting in 3{\bf k}, 5{\bf k}, and higher-order harmonic reflections \cite{Artyukhin, Roessli, Choi}.
In fact, the observed behaviour is rather reminiscent of domain patterns observed in two-component systems \cite{Seul}.

\section{Discussion}

To thoroughly address the novel stripe phase, we first need to understand the two adjacent phases.
The zero-field VC ground state is defined by {\bf k}, which complies with the spin-1/2 zigzag chain model \cite{Sudan}, yielding $J_1/J_2$\,=\,$-1.25$.
The same model accounts for the collinear SDW phase at higher magnetic fields or temperature, exhibiting a weak field and temperature dependence of the main magnetic reflection, {\bf k}($T, B$).
On the contrary, the nanometer-scale modulation in the stripe phase is far from being parallel to the main {\bf k} modulation (Fig.\,\ref{mag-struc}b-\ref{mag-struc}e), i.e., $\Delta${\bf k}\,$\cdot$\,{\bf k}\,$\ne$\,0, hence, the stripes cannot be explained by the intrachain interactions alone.

The role of the weak interchain interactions $J'$ (see Methods) in the stripe phase must be related to the competition between the VC and SDW phases.
Namely, with increasing magnetic field, the energy difference between these phases decreases and at $B$\,$\sim$\,$B_{\text{c1}}$ it becomes comparable to the strength of $J'$.
Similarly, the increasing temperature leads to the same kind of phase competition near $T_{\text{N3}}$; its effect on the energy difference between the two states reflects in the temperature dependent {\bf k}($T$) (Fig.\,\ref{ND}d).
In the case of non-frustrated $J'$, $|k_z/k_x|$\,$\equiv$\,2 would be independent of the magnetic field or temperature.
This is contradicted by the experiment, as the ratio $|k_z/k_x|$ noticeably changes from 2.11 to 2.04 between $T_{\text{N1}}$ and $T_{\text{N3}}$, respectively (Fig.\,\ref{ND}).
Therefore, $J'$ must be frustrated and is thus most probably essential for the nanometer-scale stripe modulation  (Fig.\,\ref{mag-struc}b-\ref{mag-struc}e) that releases the degeneracy of the VC and SDW states.

Moreover, the observed stripes are in phase, as the satellite reflections appearing around a sharp central magnetic peak are also sharp (Inset to Fig.\,\ref{ND}a). Other presently known spin stripe systems, e.g., ferromagnetic stripe domains \cite{Giuliani} and spin stripes in high-temperature superconductors \cite{Vojta}, have antiphase domain boundaries, which reflect long-range repulsive forces.
This difference implies that a conceptually different coupling mechanism must be responsible for the stripe formation in $\beta$-TeVO$_4$.

Indeed, the observed stripe state is antiferromagnetic and therefore lacks magnetized domains that could mediate sizable long-range magnetic (dipole-dipole) interactions.
Since the stripe phase is sandwiched between the VC and SDW phase, the microscopic mechanism for the stripe formation is likely the coupling of two corresponding order parameters, in the analogy to biological-membrane systems \cite{Andelman, Seifert, Baumgart}.
In this case, the stripe formation would rely on the competition between the magnetic entropy contribution that decreases with decreasing temperature and thus favours fully developed moments of the VC phase, and the magnetic anisotropy energy, which in an uniaxial case favours collinear magnetic moments of the SDW phase.
This seems to be in line also with finite, though very weak, chiral terms in the intermediate phase (Table\,\ref{tabSNP}) and may also lead to asymmetric intensity of satellite reflections (Inset in Fig.\,\ref{ND}a).
The stripe formation may be further assisted by the possible electric polarization, since the latter is allowed in the VC phase due to symmetry reasons as argued in magnetoelectric multiferroics \cite{Wang}.
Such a novel state may be inherent to quasi one-dimensional chiral multiferroics, as stripes of spiral and collinear phases have recently been found also in MnWO$_4$ \cite{Taniguchi}.

Our results exceed the limits of standard spin-stripe models assuming that stripe patterns stem from the competition between short-range attractive and long-range repulsive magnetic forces  \cite{DeBell, Portmann, Mu, Giuliani}.
We demonstrate that $\beta$-TeVO$_4$, which has proven to be a nearly perfect realization of the highly-frustrated ferromagnetic spin-1/2 chain, is a model system for spin-stripe formation in the absence of sizable long-range magnetic interactions.
The dominant intrachain couplings are accompanied by weak frustrated short-range interchain exchange interactions and possibly also by electric polarization, which stabilize a nanometer-scale modulation of a purely antiferromagnetic order.
The proposed concept thus reveals a new perspective on the stripe pattern formation in strongly correlated electron systems and may prove valuable for understanding of other, yet unexplained, phenomena, e.g., the charge modulation in high-temperature superconductors and its relation to superconductivity.
Moreover, the frustrated nature of $J'$ makes this system highly susceptible to external perturbations and, therefore, allows for a direct manipulation of the nanometer-scale modulation by either changing the temperature or the strength of the applied magnetic (and possibly also electric) field.
Finally, understanding the origin of the stripe phase may prove essential in the search for elusive spin-nematic phase that is predicted close to the saturation field, which is in $\beta$-TeVO$_4$ conveniently low in respect to other candidate systems \cite{Svistov}.

\section{Methods}

{\bf Sample description.}
All measurements were conducted on the same high-quality single-crystal samples.
These were obtained from TeO$_2$ and VO$_2$ powders by chemical vapor transport reaction, using two-zone furnace and TeCl$_4$ as a transport agent. Detailed reaction conditions are reported in Ref. \cite{Meunier1}.

{\bf Neutron diffraction.}
Neutron experiments on a 2$\times$3$\times$4\,mm$^3$ single crystal were performed on TriCS diffractometer at the Paul Scherrer Institute (PSI), Switzerland.
The zero-field diffraction data between 1.7 and 5\,K were collected using a cooling machine in a four-circle geometry.
Magnetic field dependences were measured in a vertical magnet in the normal beam (tilt) geometry.

The monoclinic unit cell  (space group $P$2$_1$/$c$) with parameters 
$a$\,=\,4.3919(1)\AA, $b$\,=\,13.5155(1)\AA, $c$\,=\,5.4632(1)\AA\, and $\beta$\,=\,90.779(1)$^\circ$ was determined from single-crystal neutron diffraction data (440 reflections) collected at 10\,K. 
These data were used also to refine the atomic positions (Table\,\ref{tab1}) and yield the reliability factor $R_{{\rm obs}}$\,=\,9.03.

We note that width of nuclear and magnetic peaks measured in transversal to {\bf k} (omega-scans) is comparable and is very close to the instrumental resolution.
The {\bf k}-scan presented in the inset in Fig.\,\ref{ND}a was performed in a 4-circle geometry.
The peak shape is dominated by instrumental resolution (with vertical component rather relaxed) and is not intrinsic to the crystal.

The scan in Fig.\,\ref{ND}b extended from (0.204~2.05~0.558) at $\omega$\,=\,$-$107.5$^\circ$ to (0.211~1.965~0.598) at $\omega$\,=\,$-$104.5$^\circ$.

\begin{table} [!]
\caption{{\bf Fractional atomic coordinates. The values derived} for $\beta$-TeVO$_4$ at 10\,K (see \cite{Jana}).
\label{tab1}}
\begin{ruledtabular}
\begin{tabular}{cccc} 
atom  &  x  &  y  &  z \\
\hline
Te& 0.0417(9)& 0.3910(1)& 0.6405(8) \\
V& 0.666(11)&0.158(3)& 0.639(11) \\
O1& 0.3047(9)& 0.1646(2)& 0.6683(9) \\
O2& 0.8294(9)& 0.0473(2)& 0.8664(9) \\
O3& 0.8062(9)& 0.2221(2)& 0.9806(9) \\
O4& 0.7467(9)& 0.0812(2)& 0.3700(9) \\
\end{tabular}
\end{ruledtabular}
\end{table}

\begin{table} [!]
\caption{{\bf Interatomic distances.} The values are given in \AA\, for $\beta$-TeVO$_4$ at 10\,K for the most probable interchain exchange interactions ({\bf 1}: V-O1-Te-O2-V and {\bf 2}: V-O1-Te-O4-V), (see \cite{Jana}).
\label{tab2}}
\begin{ruledtabular}
\begin{tabular}{ccccc}
path & $d_{\rm{V-O}}$ & $d_{\rm{O-Te}}$ & $d_{\rm{Te-O}}$ & $d_{\rm{O-V}}$ \\
\hline
{\bf 1} & 1.60(5)    & 2.939(19)    & 1.940(12)   & 2.06(5) \\
{\bf 2} & 1.60(5)    & 2.939(19)    & 1.854(12)   & 1.83(5)\\

\end{tabular}
\end{ruledtabular}
\end{table}

{\bf Spherical neutron polarimetry.}
Experiments on a 2$\times$3$\times$4\,mm$^3$ single crystal were performed on D3 diffractometer equipped with CRYOPAD operating with wavelength $\lambda$\,=\,0.825\,\AA\, at the Institute Laue Langevin (ILL), Grenoble, France.
The low-temperature diffraction data between 1.7 and 5\,K were collected using ILL Orange cryostat.
The two different scattering planes, namely ($h$00)/(00$l$) and ($-h$02$h$)/(0$k$0), were accessed by using two different mountings of the crystal.

{\bf Characterization of the exchange pathways.}
Derived atomic positions provide an additional insight into the nature of the dominant exchange interactions.
In particular, the ferromagnetic nature of $J_1$ complies with the dependence of the exchange coupling on the distance between the V$^{4+}$ ion and the basal oxygen plane of the VO$_5$ square pyramid, $\delta$.
Namely, for small $\delta$ the exchange is ferromagnetic due to the $\pi$-interaction between vanadium 3$d_{xy}$ and the basal oxygen 2$p$ orbitals, as for instance also found in CdVO$_3$ where $\delta$\,=\,0.430\,\AA \cite{Dai}.
In $\beta$-TeVO$_4$ we find reasonably small $\delta$\,=\,0.58(5)\,\AA\, at 10\,K.

On the contrary, the antiferromagnetic character of $J_2$ and $J'$ is in agreement with the findings for O$-$Te$-$O exchange bridges in oxyhalide tellurites \cite{FTOBAFMR}.
The derived $J_2$\,$\gg$\,$J'$ reflects substantially different lengths of intra- and inter-chain O$-$Te bonds that equal  1.940(12)\,\AA\, and 2.939(19)\,\AA, respectively.
Finally, assuming that the interchain $J'$ exchange interactions pass through the shortest O$-$Te interchain distance $d_{\rm{Te-O}}$\,=\,2.940(12)\,\AA , we find two similar V$-$O$-$Te$-$O$-$V paths, involving comparable intrachain O$-$Te and O$-$V distances (Table\,\ref{tab2}).

{\bf Magnetic susceptibility.}
The magnetic susceptibility $\chi(T)$ for the magnetic field {\bf B} applied along the $a$ axis was measured between 2 and 300\,K on a Quantum Design physical property measurement system (PPMS).
\begin{figure} [!] \center
\includegraphics[width=0.5\textwidth]{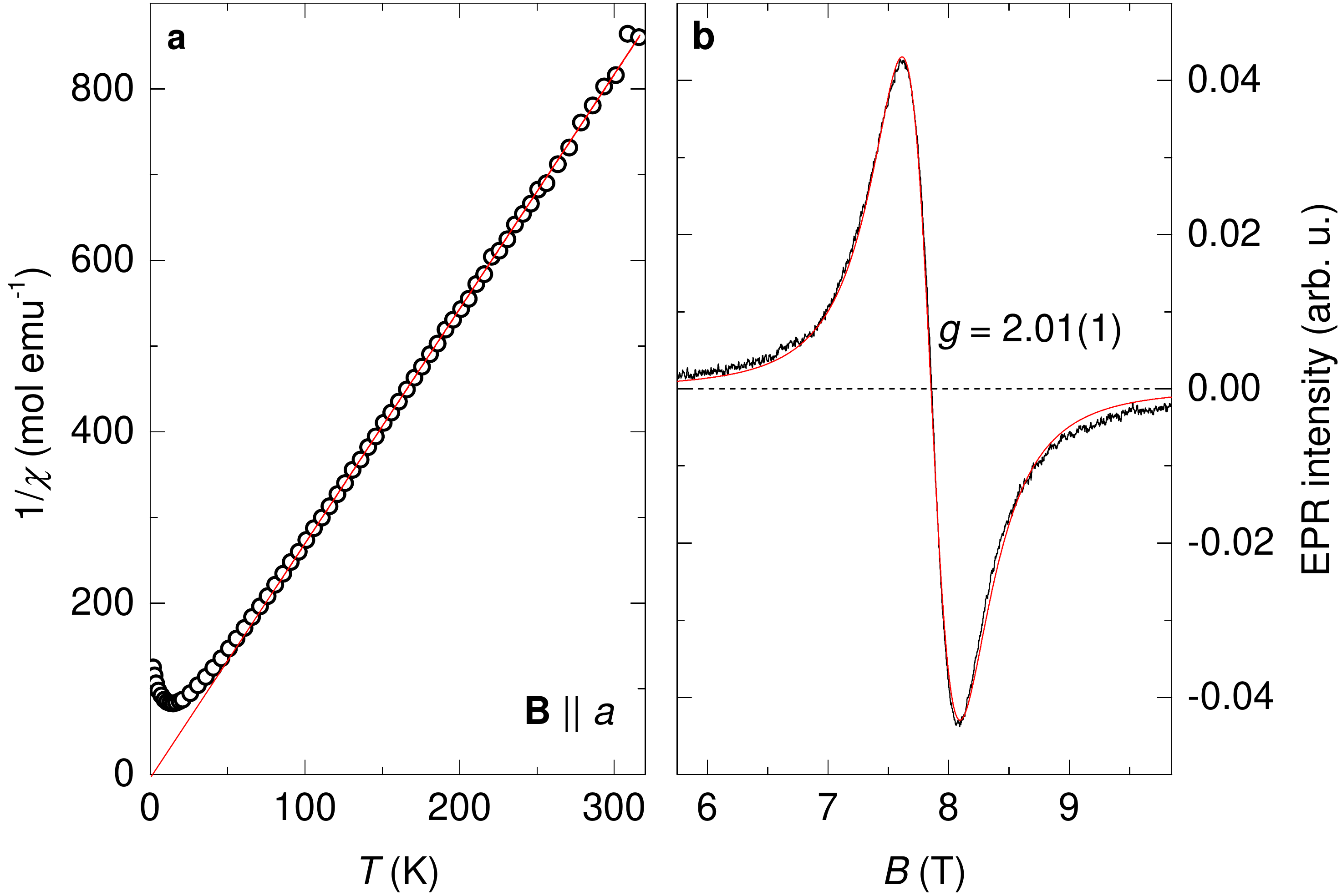}
\caption{{\bf Characterization of the paramagnetic species.} ({\bf a}) The inverse of the magnetic susceptibility in the magnetic field of 0.1\,T along the $a$ axis as a function of temperature (symbols). The solid line represents a linear fit for $T$\,$>$\,80\,K. ({\bf b}) Room-temperature electron paramagnetic resonance spectrum at 220.8\,GHz (black line), and the corresponding fit to the Lorentzian lineshape (red line), yielding the $g$ factor of V$^{4+}$ in $\beta$-TeVO$_4$.}
\label{supsuscep}
\end{figure}
The Curie-Weiss temperature $\theta$ was determined from the linear fit of the inverse susceptibility, 1/$\chi$, above 80\,K (Fig.\,\ref{supsuscep}a), measured in the magnetic field of 0.1\,T.
The temperature-independent diamagnetic contribution of the core shells is of the order of 10$^{-5}$\,emu mol$^{-1}$ and is thus negligible when compared to other terms.

{\bf High-field magnetization measurements.}
Magnetization measurements in pulsed magnetic fields up to 25\,T were performed at the High Magnetic Field Laboratory, Institute for Materials Research, Sendai.

{\bf Specific heat}.
The specific-heat measurements were performed at 0 and 3\,T between 1.8 and 30\,K using the PPMS.
The crystal-lattice contribution to the specific heat was modelled with the Debye approximation \cite{Debye}
\begin{equation}\label{clatt} 
c_{\rm{latt}}(T)=9Nk_\text{B}(T/\theta_\text{D})^3\int_0^{\theta_\text{D}/T}x^4e^x/(e^x-1)^2dx,
\end{equation}
where $\theta_\text{D}$ is the Debye temperature, $N$ is the number of atoms in the crystal, and $k_\text{B}$ is the Boltzmann constant.
The best agreement with the experiment was obtained for $\theta_\text{D}$\,=\,190\,K.

{\bf Electron paramagnetic resonance.}
Electron paramagnetic resonance (EPR) signal was measured at room temperature and 220.8\,GHz.
The experiment was conducted  on a transmission-mode EPR spectrometer at the National High Field Laboratory, Tallhassee, Florida.
The line shape  is Lorentzian and thus characteristic of exchange narrowing, as evident from the fit shown in Fig.\,\ref{supsuscep}b, which yields $g$\,=\,2.01(1) and the width $\Delta B$\,=\,0.830(2)\,T.

\section{Acknowledgments}
We acknowledge the financial support of the Slovenian Research Agency (projects Z1-5443, Bi-US/14-15-39) and the Swiss National Science Foundation (project SCOPES IZ73Z0\_152734/1).
This research has received funding from the European Union's Seventh Framework Programme for research, technological development and demonstration under the NMI3-II Grant number 283883.
The magnetic susceptibility and specific heat measurements were measured in the Laboratory for Developments and Methods, Paul Scherrer Institute, Villigen, Switzerland, while neutron diffraction experiments were performed at the Swiss spallation neutron source SINQ, at the same institute and at the reactor of the Institute Laue-Langevin, Grenoble, France.
The access to the NHMFL and the technical assistance of A. Ozarowski with the EPR measurements is acknowledged.
H.N. acknowledges ICC-IMR and KAKENHI23224009.\\

\section{Author contributions}

M. P., A. Z., O. Z., and D. A. designed and supervised the project.
The samples were synthesized by H. B.
The neutron diffraction experiments were performed and analyzed by O. Z. and M. P.
The spherical neutron polarization experiments were performed and analyzed by O. Z., M. P. and L. C. C.
The magnetic susceptibility and specific heat measurements were conducted by O. Z. and analyzed by M. P.
The high field magnetization measurements were performed by H. N. and M. P. and analyzed by M. P.
All authors contributed to the interpretation of the data and to the writing of the manuscript.

\section{Additional information}
{\bf Competing financial interest:} The authors declare no competing financial interests.


\begin{thebibliography}{99}



\bibitem{Seul} Seul, M. and Andelman, D. {\rm  Domain shapes and patterns: The phenomenology of modulated phases.} {\em Science} {\bf 267,} 476-483 (1995). 
\bibitem{Stoycheva} Stoycheva, A. D. and Singer, S. J. {\rm Stripe melting in a two-dimensional system with competing interactions.} {\em Phys. Rev. Lett.} {\bf 84,} 4657-4660 (2000).
\bibitem{Malescio} Malescio, G. and Pellicane, G. {\rm Stripe phases from isotropic repulsive interactions.} {\em Nat. Mater.} {\bf 2,} 97-100 (2003).

\bibitem{DeBell}  De'Bell, K.,  MacIsaac, A. B., and  Whitehead, J. P. {\rm Dipolar effects in magnetic thin films and quasi-two-dimensional systems.} {\em Rev. Mod. Phys.} {\bf 72,} 225-257 (2000).
\bibitem{Portmann}  Portmann, O.,  Vaterlaus, A., and  Pescia, D. {\rm An inverse transition of magnetic domain patterns in ultrathin films.} {\em Nature} {\bf 422,} 701-704 (2003).
\bibitem{Mu}  Mu, Y. and  Ma, Y.  {\rm Stripe patterns in frustrated spin systems.} {\em J. Chem. Phys.} {\bf 117,} 1686 (2002).
\bibitem{Giuliani}  Giuliani, A., Lebowitz, J. L., and  Lieb, E. H. {\rm Checkerboards, stripes, and corner energies in spin models with competing interactions.} {\em Phys. Rev. B} {\bf 84,} 064205 (2011).


\bibitem{Andelman} Andelman, D., Kawakatsu, T., and Kawasaki, K. {\rm Equilibrium shape of two-component unilamellar membranes and vesicles.} {\em Europhys. Lett.} {\bf 19,} 57-62 (1992).
\bibitem{Seifert} Seifert, U. {\rm Configurations of fluid membranes and vesicles.} {\em Adv. Phys.} {\bf 46,} 13-137 (1997).
\bibitem{Baumgart} Baumgart, T., Hess, S. T., and Webb, W. W., {\rm Imaging coexisting fluid domains in biomembrane models coupling curvature and line tension.} {\em Nature} {\bf 425,} 821-824 (2003).

\bibitem{Tranquada} Tranquada, J. M., Sternlieb, B. J., Axe, J. D., Nakamura, Y., and Uchida, S. {\rm Evidence for stripe correlations of spins and holes in copper oxide superconductors.} {\em Nature} {\bf 375,} 561-563 (1995).
\bibitem{Emery} Emery, V. J., Kivelson, S. A., and Tranquada, J. M. {\rm Stripe phases in high-temperature superconductors.} {\em Proc. Natl. Acad. Sci. USA} {\bf 96,} 8814-8817 (1999).
\bibitem{Vojta} Vojta, M. {\rm Lattice symmetry breaking in cuprate superconductors: stripes, nematics, and superconductivity.} {\em Adv. Phys.} {\bf 58}, 699-820 (2009).
\bibitem{Ghiringhelli} Ghiringhelli, G. {\em et al.} {\rm Long-range incommensurate charge fluctuations in (Y,Nd)Ba$_2$Cu$_3$O$_{6+x}$.} {\em Science} {\bf 337,} 821-825 (2012).
\bibitem{Wu2013} Wu, T., {\em et al.} {\rm Emergence of charge order from the vortex state of a high-temperature superconductor.} {\em Nat. Comm.} {\bf 4,} 2113 (2013).






\bibitem{Sudan} Sudan, J., L\"{u}scher, A., and L\"{a}uchli, A. M. {\rm Emergent multipolar spin correlations in a fluctuating spiral: The frustrated ferromagnetic spin-1/2 Heisenberg chain in a magnetic field.} {\em Phys. Rev. B} {\bf 80,} 140402(R) (2009).
\bibitem{Hikihara} Hikihara, T., Kecke, L., Momoi, T., and Furusaki, A. {\rm Vector chiral and multipolar orders in the spin-1/2 frustrated ferromagnetic chain in magnetic field.} {\em Phys. Rev. B} {\bf 78,} 144404 (2008).
\bibitem{Hikihara2010} Hikihara, T., Momoi, T., Furusaki, A., and Kawamura, H. {\rm Magnetic phase diagram of the spin-1/2 antiferromagnetic zigzag ladder.} {\em Phys. Rev. B} {\bf 81,} 224433 (2010).
\bibitem{Okunishi} Okunishi, K. and Tonegawa, T. {\rm Magnetic phase diagram of the $S$\,=\,1/2 antiferromagnetic zigzag spin chain in the strongly frustrated region: cusp and plateau.} {\em J. Phys. Soc. Jpn.} {\bf 72,} 479-482 (2003).
\bibitem{McCulloch} McCulloch, I. P., Kube, R., Kurz, M., Kleine, A., Schollw\"{o}ck, U., and Kolezhuk, A. K. {\rm Vector chiral order in frustrated spin chains.} {\em Phys. Rev. B} {\bf 77,} 094404 (2008).





\bibitem{Schapers} Sch\"{a}pers, M. {\em et al.} {\rm Thermodynamic properties of the anisotropic frustrated spin-chain compound linarite PbCuSO$_4$(OH)$_2$.} {\em Phys. Rev. B} {\bf 88,} 184410 (2013) and references therein.
\bibitem{Sato13} Sato, M., Hikihara, T., and Momoi, T. {\rm Spin-nematic and spin-density-wave orders in spatially anisotropic frustrated magnets in a magnetic field.} {\em Phys. Rev. Lett.} {\bf 110,} 077206 (2013).

\bibitem{Svistov} Svistov, L. E. {\rm New high magnetic field phase of the frustrated $S$\,=\,1/2 chain compound LiCuVO$_4$.} {\em J. Exp. Theor. Phys. Lett.} {\bf 93,} 21-25 (2011).

\bibitem{Chakrabarty} Chakrabarty, T., Mahajan, A. V., Gippius, A. A., Tkachev, A. V., B\"{u}ttgen, N., and Kraetschmer, W. {\rm BaV$_3$O$_8$: A possible Majumdar-Ghosh system with $S$\,=\,1/2.} {\em Phys. Rev. B} {\bf 88,} 014433 (2013).
\bibitem{Kataev} Kataev, V. {\rm Interplay between structure, transport and magnetism in the frustrated $S$\,=\,1/2 system In$_2$VO$_5$.} {\em J. Phys. Conf. Ser.} {\bf 150,} 042084 (2009).

\bibitem{Meunier2} Meunier G., Darriet, J., and Galy, J. {\rm L'Oxyde double TeVO$_4$ II. Structure cristalline de TeVO$_4$-$\beta$-relations structurales.} {\em J. Sol. Stat. Chem.} {\bf 6,} 67-73 (1973).
\bibitem{Savina} Savina, Y., Bludov, O., Pashchenko, V., Gnatchenko, S. L., Lemmens, P., and Berger, H. {\rm Magnetic properties of the antiferromagnetic spin-1/2 chain system $\beta$-TeVO$_4$.} {\em Phys. Rev. B} {\bf 84,} 104447 (2011).

\bibitem{Bonner} Bonner, J. and Fisher, M. {\rm Linear Magnetic Chains with Anisotropic Coupling.} {\em Phys. Rev.} {\bf 135,} A640-A658 (1964).

\bibitem{Debye} Debye, P. {\rm Zur Theorie der spezifischen Waerme.} {\em Ann. Phys. (Leipzig)} {\bf 39,} 789-839 (1912).


\bibitem{Tsirlin} Tsirlin, A. A., Janson, O., and Rosner, H. {\rm Unusual ferromagnetic superexchange in CdVO$_3$: The role of Cd.} {\em Phys. Rev. B} {\bf 84,} 144429 (2011).
\bibitem{Dai} Dai, D., Koo, H.-J., and Whangbo, M.-H. {\rm Analysis of the spin exchange interactions of ferromagnetic CdVO$_3$ in terms of first principles and qualitative electronic structure calculations.} {\em J. Solid State Chem.} {\bf 175,} 341-347 (2003).
\bibitem{FTOBAFMR} Pregelj, M. {\em et al.} {\rm Multiferroic FeTe$_2$O$_5$Br: Alternating spin chains with frustrated interchain interactions.} {\em Phys. Rev. B} {\bf 86,} 054402 (2012).


\bibitem{Bursill} Bursill, R., Gehring, G. A., Farnell, D. J. J., Parkinson, J. B., Xiang, T., and Zeng, C. {\rm Numerical and approximate analytical results for the frustrated spin-1/2 quantum spin chain.} {\em J. Phys.: Condens. Matter} {\bf 7,} 8605-8618 (1995).
\bibitem{White} White, S. R., and Affleck, I. {\rm Dimerization and incommensurate spiral spin correlations in the zigzag spin chain: Analogies to the Kondo lattice.} {\em Phys. Rev. B} {\bf 54,} 9862-9869 (1996).

\bibitem{Koo} Koo, H.-J. and Whangbo, M.-H. {\rm Analysis of the spin–spin interactions in layered oxides $\alpha'$-NaV$_2$O$_5$, CaV$_2$O$_5$ and MgV$_2$O$_5$ and the spin-Peierls distortion in $\alpha'$-NaV$_2$O$_5$ by molecular orbital, Madelung energy and bond valence sum calculations.} {\em Solid State Commun.} {\bf 111,} 353-360 (1999).
\bibitem{Onoda} Onoda, M. and Nishiguchi, N. {\rm $S$\,=\,1/2 zigzag-chain structure and ferromagnetism of CdVO$_3$.} {\em J. Phys.: Condens. Matter} {\bf 11,} 749-757 (1999).
\bibitem{Pickett} Pickett, W. E. {\rm Impact of Structure on Magnetic Coupling in CaV$_4$O$_9$.} {\em Phys. Rev. Lett.} {\bf 79,} 1746-1749 (1997).






\bibitem{Heidrich-Meisner} Heidrich-Meisner, F., Honecker, A., and Vekua, T. {\rm Frustrated ferromagnetic spin-1/2 chain in a magnetic field: The phase diagram and thermodynamic properties.} {\em Phys. Rev. B} {\bf 74,} 020403(R) (2006)
\bibitem{Heidrich-Meisner-online} Honecker, A. {\rm Thermodynamics of the $J_1$-$J_2$ spin-1/2 Heisenberg chain.} http://www.theorie.physik.uni-goettingen.de/~honecker/j1j2-td/ (2006).

\bibitem{Carlin} Carlin, R. L. {\em Magnetochemistry} (Springer-Verlag, Berlin, 1986).

\bibitem{Kittel} Kittel, C. {\em Introduction to solid state physics} 8th edn (John Wiley \& sons, New York, 2005).


\bibitem{Artyukhin} Artyukhin, S. {\em et al.} {\rm Solitonic lattice and Yukawa forces in the rare-earth orthoferrite TbFeO$_3$.} {\em Nature Mater.} {\bf 11,} 694-699 (2012).
\bibitem{Roessli} Roessli, B. {\em et al.} {\rm Formation of a magnetic soliton lattice in copper metaborate.} {\em Phys. Rev. Lett.} {\bf 86,} 1885-1888 (2001).
\bibitem{Choi} Choi, S.-M., Lynn, J.W., Lopez, D., Gammel, P. L., Canfield, P. C., and Bud’ko,  S. L. {\rm  Direct observation of spontaneous weak ferromagnetism in the superconductor ErNi$_2$B$_2$C.} {\em Phys. Rev. Lett.} {\bf 87,} 107001 (2001).

\bibitem{Wang} Wang, K.F., Liu, J. M., and Ren Z. F. {\rm Multiferroicity: the coupling between magnetic and polarization orders.} {\em Adv. Phys.} {\bf 58}, 321-448 (2009).

\bibitem{Taniguchi} Taniguchi, K., Saito, M., and Arima, T. H. {\rm Optical imaging of coexisting collinear and spiral spin phases in the magnetoelectric multiferroic MnWO$_4$.} {\em Phys. Rev. B} {\bf 81,} 064406 (2010).

\bibitem{Meunier1} Meunier G., Darriet, J., and Galy, J. {\rm L'Oxyde double TeVO$_4$ I. Synth\`{e}se et polymorphisme, structure cristalline de $\alpha$-TeVO$_4$} {\em J. Sol. Stat. Chem.} {\bf 5,} 314-320 (1972).


\bibitem{Jana} Petricek, V., Dusek, M., and Palatinus, L. {\rm Crystallographic Computing System JANA2006: General features.} {\em  Z. Kristallogr.} {\bf 229}, 345-352 (2014).






\end{thebibliography}
\end{document}